\documentclass[fleqn,twoside]{article}
\usepackage{espcrc2}

\usepackage{graphicx,amssymb,latexsym,amsmath}
\usepackage[figuresright]{rotating}

\newcommand{\beq}{\begin{eqnarray}}
\newcommand{\eeq}{\end{eqnarray}}

\title{$SU(N)$ Gauge Theories Near $T_c$\thanks{Talk presented by B.~Lucini.}}

\author{B. Lucini\address[Oxf]{Theoretical Physics, 
        Oxford University, \\ 
        1 Keble Road, Oxford OX1 3NP, United Kingdom}%
        \thanks{EU Marie Curie fellow.},
        M. Teper\addressmark[Oxf],
        U. Wenger\addressmark[Oxf]\thanks{PPARC SPG fellow.}}
       
\begin{document}

\begin{abstract}
We study the deconfinement phase transition in $SU(N)$ gauge
theories for $N$=2,3,4,6,8. The transition is first order for $N \ge 3$, 
with the strength increasing as $N$ increases. We extrapolate 
$T_c/\sqrt{\sigma}$ to the continuum limit for each $N$, and observe
a rapid approach to the large $N$ limit. As $N$ increases the phase
transition becomes clear-cut on smaller spatial volumes,
indicating the absence of (non-singular) finite volume
corrections at $N=\infty$ -- reminiscent of large $N$ reduction.
The observed rapid increase of the inter-phase surface tension  
with $N$ may indicate that for $N=\infty$ the 
deconfinement transition cannot, in practise, occur.
\vspace{1pc}
\end{abstract}

\maketitle

\section{INTRODUCTION}
\label{sec:introduction}
Originally proposed as a possible way of solving QCD \cite{thooft},
the large $N$ limit of $SU(N)$ gauge theories plays an important role for our
understanding of this class of theories.\\
It has been shown numerically \cite{blmt1} that several quantities at zero
temperature have a well-defined large $N$ limit and that deviations from that
limit at finite $N$ are small and accounted for by a ${\cal O}(1/N^2)$
correction (the first expected correction by perturbative arguments)
all the way down to $N=2$.
In \cite{blmtuw1} similar features were observed at finite temperature.
In particular, for the deconfinement temperature $T_c$ it was found 
\beq
\label{fitold}
T_c/\sqrt{\sigma} =  0.582(15) + 0.43(13)/N^2 \ .
\eeq 
Here we extend the investigation of \cite{blmtuw1}. We study $SU(N)$
groups for $N=2,3,4,6,8$. New improved results for both $T_c$ and
$\sqrt{\sigma}$ help us to keep under better control the
extrapolation to infinite $N$.\\
In the following we will discuss our numerical results for the critical
temperature and the latent heat $L_h$ as a function of $N$ and
what the observed behaviour at finite $N$ may imply for the physics
at finite temperature in the large $N$ limit. This work is based
on \cite{blmtuw2}, to which we refer the reader for a more detailed
discussion.

\section{LATTICE SETUP}
\label{sect:lattice}
In our numerical simulations we have used the Wilson action
\begin{equation}
S = \beta \sum_{p}
\left\{1-\frac{1}{N}{\mathrm {ReTr}} U_P\right\}\ ,
\label{B1}
\end{equation}
where $\beta = 2 N /g^2$, with $g$ the coupling of the theory,
$U_P$ is the path ordered product of the links around the plaquette $P$
and the links are $SU(N)$ matrices. The sum is performed over all the
plaquettes $P$. The lattice size is $L$ in three
directions and $N_t$ (with $N_t \ll L$) in the fourth one. We call spatial
directions the three of equal size and temporal direction the remaining one.\\
The system can be regarded either as a statistical system in four dimensions
with the role of temperature played by $\beta$ or as a three-dimensional
quantum field theory at finite temperature, with the physical temperature
given by
\beq
T_c = 1/N_t a(\beta) \ ,
\eeq
$a$ being the lattice spacing. Thermodynamical quantities have been
obtained via a finite size study in the
critical region of susceptibilities like the specific heat and the
susceptibility of the Polyakov loop.

\section{CRITICAL TEMPERATURE}
\label{sect:temperature}
\begin{figure}[t]
\begin{center}
\includegraphics[scale=0.35,angle=270]{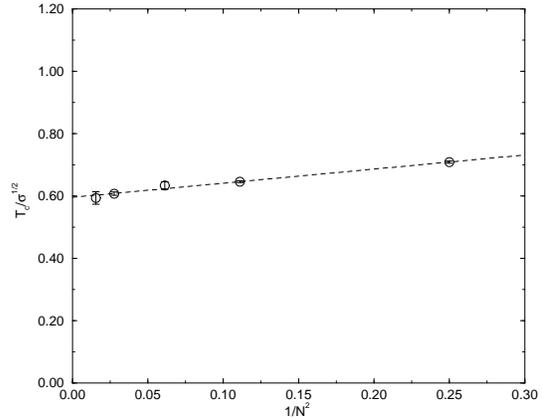}
\end{center}
\caption{Extrapolation of $T_c/\sqrt{\sigma}$ to infinite $N$. The dashed line
is the best fit to the data.}
\label{fig1}
\end{figure}

The critical temperature is unambiguously defined only in the thermodynamic
limit. On a finite lattice, one usually takes as a definition the value
at which the susceptibility of some observable relevant for the transition
has a peak. More specifically, one defines the pseudocritical coupling
$\beta_c(N_t,L)$ as the $\beta$ at which the chosen observable displays a
peak at fixed $L$, $N_t$. Then the value of the critical
coupling is given by a fit to the finite size scaling relation
\beq
\label{fitbetafss}
\beta_c(N_t,L=\infty) = \beta_c(N_t,L) + h L^{-1/\nu} \ ,
\eeq 
where, if $d$ is the number of spatial dimensions, we get $\nu = 1/d$
for a first order phase transition and $\nu < 1/d$ for a second order one.
The value of $\beta_c(N_t,L=\infty)$ does not depend on the chosen
observable.\\
In principle the order of the phase transition can be extracted from the
very same procedure used for determining $\beta_c(L=\infty)$.
However, in practise that proves to be hard: using (\ref{fitbetafss})
as a fit ansatz gives large errors on $\nu$. For this reason,
we have determined this exponent from the scaling of the maximum of
a susceptibility. In fact, from general arguments
\beq
\label{fitmaxfss}
\chi_{max}(N_t,L=\infty) \propto L^{\gamma/\nu} \ ,
\eeq
where $\gamma$ is the critical exponent governing the scaling of the
considered observable. The fitted value must satisfy the constraint
$0 < \gamma/\nu \le d$, with
the upper bound saturated by first order phase transitions.\\
The cleaner signal for the fit~(\ref{fitmaxfss}) is generally given by the
specific heat, from which we get evidence for a first order phase
transition for $N>2$.
Inserting this information back into eq.~(\ref{fitbetafss}), we obtain
$\beta_c(N_t,\infty)$. Next, by measuring the string tension at zero
temperature at that value of $\beta$, we get $T_c / \sqrt{\sigma}$.\\
We have determined $\beta_c(L=\infty)$ for at least three values of $N_t$,
chosen in such a way to avoid any bulk phase for $N \ge 4$.
This allows us to obtain the continuum limit of
$T_c/\sqrt{\sigma}$\footnote{Our finite size analysis
has been performed in full only at $N_t=5$. For other values of
$N_t$, we have studied one lattice size in the thermodynamic regime and then
we have used the scaling with $\beta$ of $h$ in eq.~(\ref{fitbetafss})
to determine $\beta_c$.}.\\
The final step is the extrapolation to $N = \infty$. We find
\beq 
\label{extrnew}
T_c/\sqrt{\sigma} = 0.596(4) + 0.453(30)/N^2 \ .
\eeq
Our data for $T_c / \sqrt{\sigma}$ as a function of $1/N^2$ are plotted in
fig.~\ref{fig1}. Despite the increased precision, which reflects on
a noticeable reduction of the error bars with respect to (\ref{fitold}),
still the leading correction in $1/N^2$ accurately describes the data
all the way down to $SU(2)$.
\begin{figure}[t]
\begin{center}
\includegraphics[scale=0.35,angle=270]{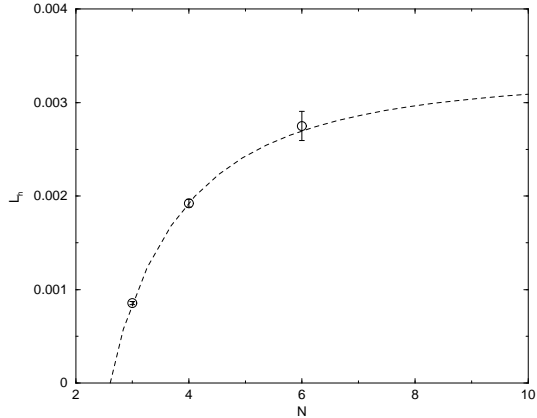}
\end{center}
\caption{Extrapolation of $L_h$ to infinite $N$. The dashed curve
is the best fit to the data.}
\label{fig2}
\end{figure}
\section{LATENT HEAT}
\label{sect:heat}
At finite $N \ge 3$ the deconfinement phase transition is first order.
First order phase transitions are characterised by
a latent heat, {\em i.e.} a difference in action between the confined
and the deconfined vacuum at $T=T_c$\footnote{Here we are referring to the
latent heat of the statistical system, which is related to the jump of the
gluon condensate across the transition in the quantum theoretical system.}.
The latent heat (per site) can be obtained from the infinite volume
limit of the maximum of the specific heat in the following way:
\beq
L_h(N_t) =  \lim_{L\to\infty} \left( \frac{2}{\beta_c(L,N_t)}
\sqrt{\frac{C_{max}(L,N_t)}{6 L^3}} \right) \ .
\eeq
We have measured $L_h$ for $N_t=5$ and $N \ge 3$, and we have found
that its value increases with $N$.
Assuming a leading ${\cal O}(1/N^2)$ correction, our results
at finite $N$ can be extrapolated to $N = \infty$ (cfr. fig.~\ref{fig2}).
We obtain a finite value for $L_h$ in that limit, which implies that the
transition is first order at $N=\infty$.
The negative coefficient for the correction indicates that the strength
of the transition increases with $N$.
\section{SURFACE AND FINITE SIZE EFFECTS}
\label{sect:surface}
A consequence of the increasing strength of the phase transition is that
a stronger interface separates the confined and the deconfined vacua
at larger $N$. Hence the life time of a metastable state at fixed
spatial volume increases as $N$ is
increased, suggesting that it becomes infinite at $N = \infty$ even for $\beta$
close to but different from $\beta_c$. In that case, we would have two
stable states in a finite region of $\beta$'s. Interpreted in terms
of the Master Field \cite{witten}, this suggests the possibility that
this field is not unique: at infinite $N$ there must be more than one
Master Field, separated from each other by infinite energy barriers.
The interplay between the different vacua at
finite but large $N$ guarantees a rich physics in the limiting case.\\ 
When we vary the volume, we observe that as $N$ increases
(a) the thermodynamical limit is reached earlier and (b) finite size
corrections are smaller. This suggests that at large $N$ the scaling
regime is obtained for $L = N_t + \epsilon$, with $\epsilon \to 0$
as $N \to \infty$. If we reduce
$L$ below $N_t$, the system undergoes a deconfinement phase transition
due to the breaking of a spatial $\mathbb{Z}(N)$ symmetry. This has been
investigated from a different perspective in \cite{kiskis}.
The breaking of a spatial  $\mathbb{Z}(N)$ symmetry could be
prevented by twisting the boundary conditions in all spatial directions. 
In this way a finite temperature regime can be obtained for $L < N_t$.
It is then plausible that at infinite $N$ the spatial volume can be shrunk to
a single plaquette. This naturally connects to the twisted Eguchi-Kawai model
\cite{TEK}.
\section{CONCLUSIONS} 
\label{sect:conclusions}
Our results for $T_c$ and $L_h$ indicates that a sensible large $N$
limit exists for $SU(N)$ gauge theories also at finite temperature.
Deviations from that limit at finite $N$ are small and can be accounted for
by a leading ${\cal O}(1/N^2)$ correction.
Our extended range of $N$ (up to eight) reveals some interesting
scenarios for the physics at $N=\infty$. At present we are extending
this analysis to other observables.

\end{document}